\documentclass[12pt,preprint]{aastex}

\def\au{{\rm AU}} 
\def\kms{{\rm km}\,{\rm s}^{-1}}

\def\pc{{\rm pc}}
\def\orb{{\rm orb}}

\begin{document}
\title{Origin of the Break and Mass-Dependence in the Wide-Binary 
Projected-Separation Distribution} 

\author{Andrew Gould and Jason Eastman}
\affil{Department of Astronomy, Ohio State University,
140 W.\ 18th Ave., Columbus, OH 43210, USA; 
gould,jdeast@astronomy.ohio-state.edu}

\begin{abstract}
The distribution of Galactic-disk
wide binaries shows a clear break in slope at projected separations of
about $r_\perp\sim 2500\,\au$ in two basically independent
surveys by Chanam\'e \& Gould and L\'epine \& Bongiorno.  The latter
also showed that the frequency of wide-binary companions to G-star
primaries declines monotonically as a function of companion mass.
We show that both effects can be explained by the operation of
Heggie's law in the typical open-cluster environments where binaries
form.  One immediate conclusion is that most Galactic-disk stars formed
in open clusters with internal dispersions of a few hundred meters
per second and disruption times of a few hundred million years.

\end{abstract}



Two separate studies demonstrate that there is break in the distribution
of Galactic-disk wide-binary 
separations at about $r_\perp\sim 2500\,\au$.  For large separations,
$\Delta\theta\ga 30''$, \citet{chaname04} find a power-law distribution
$dN/d\Delta\theta\sim \Delta\theta^{-\alpha}$, with $\alpha=1.67\pm 0.07$,
while for $10''\la\Delta\theta\la 30''$, they find a flat ``Opik's (1924) Law''
distribution, i.e., $\alpha\sim 1$.  Because the
characteristic distance of the \cite{chaname04} sample is about $d\sim 60\,\pc$,
this angular-scale break point corresponds to a physical scale of
$r_\perp\sim d\Delta\theta\sim 1800\,\au$.  \citet{lepine06} studied
wide-binaries with {\it Hipparcos} \citep{hip} primaries.  Because their
targets have parallaxes, they present their results directly in terms
of $r_\perp$ (rather than $\Delta\theta$).  They find a break at about
$r_\perp\sim 3000\,\au$.  Each estimate is uncertain at about the 20\%
level.  \citet{lepine06} are hampered by small number statistics near
the break because their bright-primary sample does not permit them to probe
inward of $\Delta\theta<20''$.  On the other hand, while \citet{chaname04}
probe to smaller angular separations and have much better statistics,
it is difficult to make precise the interpretation of their angular break 
in terms of physical scales because of the substantial uncertainties
in the individual distances of their sample.  Thus, the two studies
are broadly consistent.  Moreover, as we will show below, one actually
expects the \citet{chaname04} sample to have a somewhat smaller break point.

Neither of these two surveys can probe inward of about $r_\perp=1000\,\au$.
However, it is known from the work of \citet{dm91} that the 
G-star binary distribution roughly obeys Opik's Law over about 4.5 decades
of separation, from $10^{-1.5}\,\au$ to $10^{3}\,\au$. (The \citealt{dm91}
distribution is often expressed as a Gaussian, but it should
be remembered that the abscissa of this ``Gaussian''
is in log while the ordinate is linear.)

For binaries observed at random orientations, their semi-major axis
$a$ is related to their projected separation by 
$\langle a^{-1}\rangle =(\pi/4)\langle r_\perp^{-1}\rangle$, so the observed
break at $r_\perp\sim 2500\,\au$ corresponds to $a\sim 3000\,\au$, i.e.,
to orbital velocities $v_\orb\sim 0.5\,\kms$  (assuming a typical total
mass $M\sim 1\,M_\odot$).  

\citet{lepine06} suggested that the observed break could be due
to dynamical effects in the clusters in which the binaries were born.
Indeed, the orbital velocity at $a_{\rm break}\sim 3000\,\au$
is very similar to the
1-dimensional dispersion observed in open clusters, which typically ranges
from $\sigma\sim 0.3\,\kms$ to $\sigma\sim 1\,\kms$.  
Hence, it immediately argues
that the break is an effect of ``Heggie's (1975) Law'', which states
that ``hard binaries get harder, soft binaries get softer''.  The
boundary point of this law is defined by the internal binding energy
of the binary being equal to the mean kinetic energy of the ambient
perturbers.  Since, the binary components and the perturbers have
roughly the same mass, the break point occurs when 
$v_\orb\sim \sigma$.

Consider a ``soft binary'' population whose members are 
being injected with energy at a rate that is a function of their
binding energy $E_b$: $dE_b/dt\sim E_b^{-\beta}$.  If the population is
subjected to this process for a sufficiently long time, it will
reach a steady state with $dN/dE_b\sim E_b^\beta$, corresponding to
a separation distribution $dN/da\sim a^{-2-\beta}$.  That is,
$\alpha=2+\beta$.  To a good approximation, the injection of
energy into the binary is independent of separation, i.e., $\beta=0$.
This would predict $\alpha=2$, somewhat larger than the value 
$\alpha=1.67\pm0.07$ observed by \citet{chaname04}.  To next order,
the energy injection actually grows logarithmically with separation, 
so $\beta\ga 0$,
which goes in the wrong direction but only by a small amount.  Hence,
if the argument we are giving is correct, the explanation for the 
discrepancy in slopes must be that the cluster dissolves before
it has time to reach its ``asymptotic state''.

The timescale for binary evolution is $T\sim (n\Sigma v)^{-1}$, where
$n=\rho/m$ is the ambient perturber number density, $\rho$ is the
mass density, $m$ is the mass of a typical perturber, 
$\Sigma=\pi a_{\rm break}^2$
is the cross section for a major perturbation at the break point 
$a_{\rm break}\sim 3000\,\au$, and $v=\sigma$ is the ambient velocity.
Using the virial theorem, $4\pi G\rho \sim (\sigma/R)^2$, one finds
$T\sim R^2/(a_{\rm break}\sigma)\sim 100\,$Myr, 
where we have assumed $M\sim 2m$,
adopted a cluster radius of $R\sim 1\,\pc$, imposed Heggie's Law
($v_\orb\sim \sigma$), and dropped factors of order 
unity.  Since open clusters generally dissolve on timescales that are
one or several times this binary-disruption
timescale, it is plausible that the binary distribution does
not have time to fully reach its asymptotic state.

The explanation just given makes an important prediction.  The
binding energy of a binary scales $E_b\sim M_1 M_2/a$, where
$M_1$ and $M_2$ are the component masses.  The {\it Hipparcos} primaries
in the \citet{lepine06} study are virtually all solar-type stars, i.e.,
$M_1\sim M_\odot$.  This means that for secondaries of different masses,
the break point scales as $a_{\rm break}\sim M_2$.  It is plausible
to assume that the secondaries with $a\sim a_{\rm break}$ are initially
drawn randomly from the field population.  Then we would
predict that after the binaries diffuse to larger $a$, 
the ratio of secondaries to field stars of the same mass would fall
by a factor $a_{\rm break}^\alpha$, or in other words
as $M_2^\alpha$.  Figure 9 of \citet{lepine06} shows the frequency of
observed secondaries compared to what would be expected based on 
a distribution normalized at $4<M_V<8$, i.e., stars of mass $M_2=0.8\,M_\odot$.
The foregoing argument would predict that at $M_V=12$ 
($M_2\sim 0.25\,M_\odot$), the observed secondaries should be deficient by
a factor $\sim (0.25/0.8)^{1.67}\sim 0.15$.  The actual deficiency is
about 0.5, which is significantly less dramatic.  
Nevertheless, this figure does show the expected overall trend.
One possible explanation for the discrepancy is that the timescale
for disruption of the binaries with smaller secondaries at their
break point is considerably longer simply because their orbits present
smaller cross sections.  Since the binary disruption timescales 
at ``typical'' masses are already of order the cluster-disruption timescale, 
this increase in binary-disruption timescale could substantially mitigate the 
accelerated disruption relative to the naive scaling we have given.

The same argument predicts that the \citet{chaname04} sample should
have a smaller $a_{\rm break}$ than the \citet{lepine06} sample because
each of the latter is guaranteed to have a {\it Hipparcos} (i.e.,
roughly solar mass) component, and so to have a systematically
higher binding energy.

\begin{figure}
\plotone{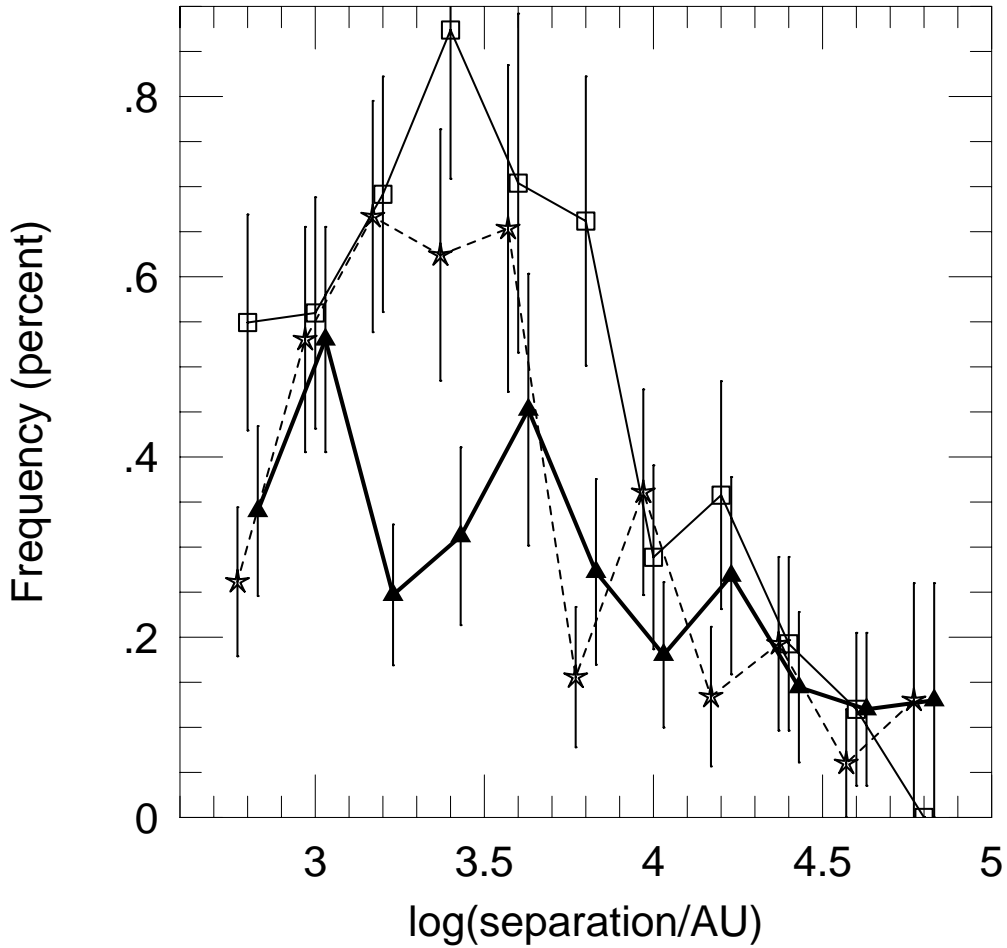}
\caption{\label{fig:f1}
Frequency of binary companions as a function of projected separation
in three luminosity bins for the secondaries: 
$M_V<8$ (triangles), $8\leq M_V < 12$ (squares),
and $M_V\geq 12$ (stars).  These correspond roughly to secondary mass
intervals $M_2>0.65\,M_\odot$, $0.25\,M_\odot<M_2<0.65\,M_\odot$,
and $M_2<0.25\,M_\odot$.  For clarity, the 
3 sets of points are slightly offset in the horizontal direction.
The data are taken form \citet{lepine06} and
correspond to their Figure 10.
According to the argument presented here,
the lower-mass bins should plateau at smaller projected separations.
The data do not show this trend but they are noisy and may still
be affected by selection.
}\end{figure}

We attempt a first test of our hypothesis by dividing the 
sample shown in \citet{lepine06} into three subsamples, with
$M_V<8$, $8\leq M_V<12$, and $M_V\geq 12$.  See Figure \ref{fig:f1}.
Our prediction would
be that the fainter stars should plateau at smaller $r_\perp$.
If there is any trend it would appear to go in the opposite
direction, although these subdivided data are quite noisy and
could still be subject to selection effects if the faintest stars
are more difficult to detect at $\sim 20''$ separations that
\citet{lepine06} believe.

An important implication of this argument is that the break in the
binary separation function tells us about the typical conditions
in which disk stars form, namely that over the lifetime of the Galactic disk,
most stars formed in clusters with
velocity dispersions of order a few hundred meters per second and that
these clusters disrupt on timescales of order a few hundred million years. 
Figure 10 of \citet{chaname04} hints that the halo
binary breakpoint occurs at at least a somewhat smaller
semi-major axis than 
that of the disk.  If this is confirmed by future observations, it would
imply that halo stars were born in environments that were kinematically
at least slightly hotter than disk stars.  (Note that it is also possible,
in principle, to produce a smaller $a_{\rm break}$ by selecting a sample
of halo binaries with systematically smaller companion masses. However,
the \citealt{chaname04} sample actually has the opposite bias: since the
halo stars are a factor $\sim 4$ farther away, they are actually biased
toward being more luminous -- and so more massive -- stars.)

Detailed verification of the scenario presented here will require
simulations that simultaneously model the disruption of the binary
and of the cluster in which it initially resides.  If verified, the
same simulations will permit more precise characterization of
the typical clusters in which today's field binaries were born.

\acknowledgments

We thank Scott Gaudi for valuable comments on the manuscript.
This work was supported by NSF grant AST 042758.
Any opinions, findings, and conclusions or recommendations expressed in
this material are those of the authors and do not necessarily reflect the
views of the NSF.

\end{document}